\documentclass[prb,preprint]{revtex4-1} % style for Physical Review B and AJP are similar

\usepackage{amsmath} % need for subequations

\usepackage{graphicx} % for figures
\usepackage{epstopdf}
\usepackage{slashed}
\usepackage{float}

\begin{document}
\raggedbottom

\title{On Back-reaction in Special Relativity}
%Lines break automatically or can be forced with \\
\author{William M. Nelson}
\email{wmn@cox.net} %optional
\date{\today}

\begin{abstract}
Back-reaction of fields plays an important role in the generation of particle masses and the mass-energy
equivalence of special relativity, 
but the most natural demonstrations using classical models result in apparent errors
such as the notorious ``4/3'' problem. Here we discuss the resolution
of these discrepancies within the underlying atomic description of matter,
with the aim of encouraging classroom discussion of back-reaction and its
connection to relativistic effects.  
 \end{abstract}
\maketitle

\section{\label{sec:intro}Introduction}

Equivalence between mass and energy is a cornerstone of special relativity, yet
the formal derivations provide little physical intuition. Students can calculate quantitatively 
correct values for the masses of different configurations, e.g. a capacitor with varying 
plate separation, but often gain little or no qualitative understanding of the mechanisms 
which produce the differences. 

One of the primary mechanisms has been understood, at least in principle,
since the 1881 observation of J.J. Thomson that a charged, accelerating sphere 
experiences a ``back-reaction'' force from its own electromagnetic (EM) field, 
which acts to resist the acceleration and effectively increases the mass of the 
sphere.\cite{thomp,griff,feyn}  This effect, and its extension to
other systems of charges such as capacitors or atomic bound states, provides the mechanism
by which internal EM energy contributes to the overall mass of a system.  The magnetic
component of the EM back-reaction should already be familiar to most
students as the phenomenon of self-induction.

Different charge configurations lead to different back-reaction forces, which, along with
internal kinetic energy such as electron orbital motion, accounts for most of the 
mass differential between different states of ordinary matter. Moreover, as has
been emphasized by Wilczek,\cite{wil1} back-reaction of the strong nuclear force accounts
for most of the mass of protons and neutrons, only a small fraction
of which comes from the intrinsic (or Higgs-derived) quark masses. 
Students who have studied self-induction therefore understand, in essence, one of the
primary sources of mass in the universe, as well as one of the principal
mechanisms of mass/energy equivalence, but they may not be aware of
these connections.  

Given the centrality of this mechanism it would be beneficial to expose
students to it, for example by computing 
back-reaction in simple classical configurations and showing its effect on mass.
However, a vexing problem arises: computations 
of this sort do not give correct answers.
One expects to find that the mass contributed by back-reaction equals 
the stored EM energy divided by $c^2$, as predicted by Einstein's formula, but
model computations give answers differing from this by
geometry-dependent factors, most notoriously the
factor 4/3 for spherically symmetric cases; indeed, the problem has come to
be known as the ``4/3 problem''.\cite{res} 

An explanation for such discrepancies was proposed long ago by Poincar\'{e},
who noted that the charge distributions appearing
in classical models are unstable and require
additional forces to stabilize them.\cite{poinc} 
Adding in the energy and momentum contributions from these ``Poincar\'{e} stresses''
compensates the discrepancy and allows the overall momentum and energy to 
transform properly as a 4-vector.\cite{res} 

Poincar\'{e} stresses do resolve the problem, but when treated generically
they leave the mechanism rather obscure, and the focus on 
additional forces is confusing since everyday matter is, in fact, constructed 
from EM forces and charged particles. 
Here we clarify how the problem is resolved within realistic atomic
matter, with the aim of encouraging classroom discussion of back-reaction 
examples. The 4/3 problem has, of course, been discussed many times previously,
but most often in the context of constructing classical models of the electron, which
is not our present interest.\cite{elec}

We view this as part of a broader effort to introduce more qualitative or constructive
explanations into the teaching of special relativity, with the goal of improving
students' intuition for relativistic effects and showing how those effects connect to prior elements
of the physics curriculum.\cite{bell, brown, miller, rmr}
No originality is claimed for the calculations to follow, 
and references have been cited when known. 
Natural units $c=\hbar=1$ will be used throughout, and ``mostly-plus''
metric signature $(-,+,+,+)$.\cite{burgess}  
Greek indices $\mu,\nu,\dots$ refer to 4-space coordinates,
while $i,j,\dots$ refer to 3-space, with repeated indices summed in both cases.

\section{\label{sec:models}Models of back-reaction and resulting discrepancies}

\begin{figure}[H]
\includegraphics{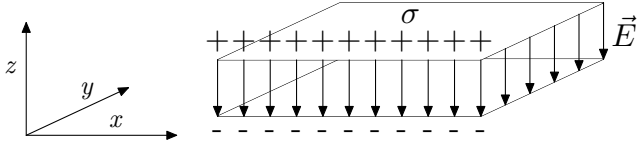}
\caption{Parallel-plate capacitor}
\label{fig:cap}
\end{figure}

Perhaps the simplest configuration in which to study EM back-reaction is the parallel-plate capacitor,
Fig.\ref{fig:cap}. This also provides a model for bound states in general,
since varying the separation of the plates changes the internal EM energy just as,
e.g., for different atomic levels.

We consider a capacitor having square plates with sides
of length $L$ parallel to the $x$, $y$ axes, separated by distance $h$ in the $z$ direction.
The top plate holds uniform surface charge density $\sigma > 0$ and the bottom plate $-\sigma$,
leading to the electric field $\vec{E} = -\sigma\hat{z}$. 

The question to be answered is how much of the mass of the capacitor 
is due to its EM field.  According to Einstein's formula the answer should be
\begin{align}
\label{expect}
\Delta M = \mathcal{H}_{EM}
\end{align}
where $\mathcal{H}_{EM}$
is the EM field energy, given by  
\begin{align}
\label{HEM}
\mathcal{H}_{EM} = {1\over 2}\int d^{3}x \,E^2.
\end{align}

One measures mass by trying to accelerate the system, i.e. 
by applying a force and measuring the total impulse required
for the system to attain a given speed.
Assuming that the external force doesn't interact with the internal EM field directly,
the only way in which the internal EM field
affects this measurement is through back-reaction onto the charges. 

We consider first accelerating the capacitor in the $\hat{x}$ direction, in which case the back-reaction 
takes the form of self-induction. As the charges accelerate, a magnetic field $\vec{B}$ develops
between the plates, parallel to $\hat{y}$. The growing B-field in turn creates
an E-field which acts back on the charges, opposing the acceleration. 
Instead of following this process explicitly, we will calculate
the total back-reaction impulse in a more general way which applies more easily
to other configurations.  

The back-reaction is just force applied by the EM field, i.e., momentum exchanged between
the field and the matter. The total back-reaction impulse generated by an 
EM field between two time points is therefore the negative of the change
in the EM field momentum. The EM field momentum is given by the Poynting vector\cite{jackson}
\begin{align}
\label{PEM}
\vec{\mathcal{P}}_{EM} = \int d^{3}x \,(\vec{E}\times\vec{B}) 
\end{align}
and we note that the capacitor has $\vec{\mathcal{P}}_{EM}=0$ when at rest. The cumulative back-reaction impulse
from accelerating the capacitor to velocity $\vec{\beta}$ is then found by calculating $\vec{\mathcal{P}}_{EM}$
for the moving system, a calculation which is often quite easy because one can use a Lorentz transformation
to find the EM field of the moving system. 

Here, and in the remainder of the paper, 
we consider only small velocities, and work to first order in $\vec{\beta}$. 
First-order calculations suffice to determine the effective mass and exhibit 
the primary back-reaction mechanisms, while simplifying the calculations greatly. 

The expected EM momentum in light of Einstein's formula (Eq.\ref{expect}) would be  
\begin{align}
\label{emlor}
\vec{\mathcal{P}}_{EM} = \vec{\beta}\mathcal{H}_{EM},
\end{align}
but the actual situation is not so simple. 
Applying the standard Lorentz transformation of the EM field,
the boosted state acquires a magnetic field $\vec{B} = \vec{\beta}\times\vec{E}$, leading to momentum\cite{griff} 
\begin{align}
\label{boostedP}
\vec{\mathcal{P}}_{EM} &= \int d^{3}x \,(\vec{E}\times(\vec{\beta}\times\vec{E}))
\nonumber\\
&= \int d^{3}x \, (\vec{\beta}E^2 - \vec{E}(\vec{\beta}\cdot\vec{E})),
\end{align}
a formula which applies to any configuration having negligible initial B-field. 

For the capacitor accelerated along $\hat{x}$, the second term vanishes, and 
by comparing with Eqs.(\ref{HEM},\ref{emlor}) one
sees immediately that the result is too large by a factor of two! 
Hence, although the example clearly shows a qualitative effect of back-reaction
on mass, it does not successfully reproduce the quantitative mass/energy relation. 

For acceleration in the $\hat{z}$ direction the situation is equally
perplexing, for no B-field is generated and the final
EM momentum is also zero.\cite{rindler} It would appear that there is no back-reaction at all, hence
no contribution from the EM energy to the mass of the system.  
One might say that the $\hat{x}$ case has a ``factor of two'' discrepancy, while the $\hat{z}$
case has a ``factor of zero'' discrepancy. In fact the discrepancy was discovered in studying
spherically-symmetric geometries (models of the electron) and has long
been referred to as the ``$4/3$ problem'', after the factor which is found for these cases. 
We note that, in addition to self-induction, one must usually also include
electrostatic self-forces; however, Eq.(\ref{boostedP}) remains valid. 

If the momentum stored in the EM field does not account for the 
relativistically-required contribution Eq.(\ref{emlor}), the difference must
be stored somewhere else, since the underlying theory of quantum
electrodynamics does respect both momentum conservation and Lorentz invariance. 
The only place additional momentum could be stored is
in the surrounding atomic material of the capacitor, hence there must
be a compensating momentum discrepancy in the atomic matter; in other words, 
the total momentum of the electrons and nuclei in the moving atomic system must
also differ from that expected by simply applying Einstein's formula (the analog
of Eq.\ref{emlor} for the matter subsystem). 
In the following sections we
show how this ``matter momentum discrepancy'' arises. 

For future reference we show the general form of the EM momentum discrepancy, 
obtained by subtracting Eq.(\ref{emlor}) from Eq.(\ref{boostedP}):
\begin{align}
\label{discP}
\vec{\Delta\mathcal{P}}_{EM} &= 
		\int d^{3}x \, ({1\over 2}\vec{\beta}E^2 - \vec{E}(\vec{\beta}\cdot\vec{E})).
\end{align}

\section{\label{sec:matter}Momentum Discrepancy in the Matter System}

If matter consisted of a static assembly of components then it would be hard to see
how an object could harbor any momentum beyond that of its bulk motion; however,
the modern view of matter involves a great deal of internal motion which can, in fact,
develop a net momentum in relativistic theories. 
We consider first a classical atomic model consisting of a uniformly distributed ring of 
identical charged particles in circular orbits
at fixed radius around a central charged nucleus. We neglect radiation, and 
assume that the total orbiting charge density is small enough that the charges
have negligible effect on each other's motion. The force on a charge $q$ located
at position $\vec{r}$ relative to the nucleus is 
$\vec{F} = -q|E|\hat{r}$, where $|E|$ is constant. 

\begin{figure}[H]
\includegraphics{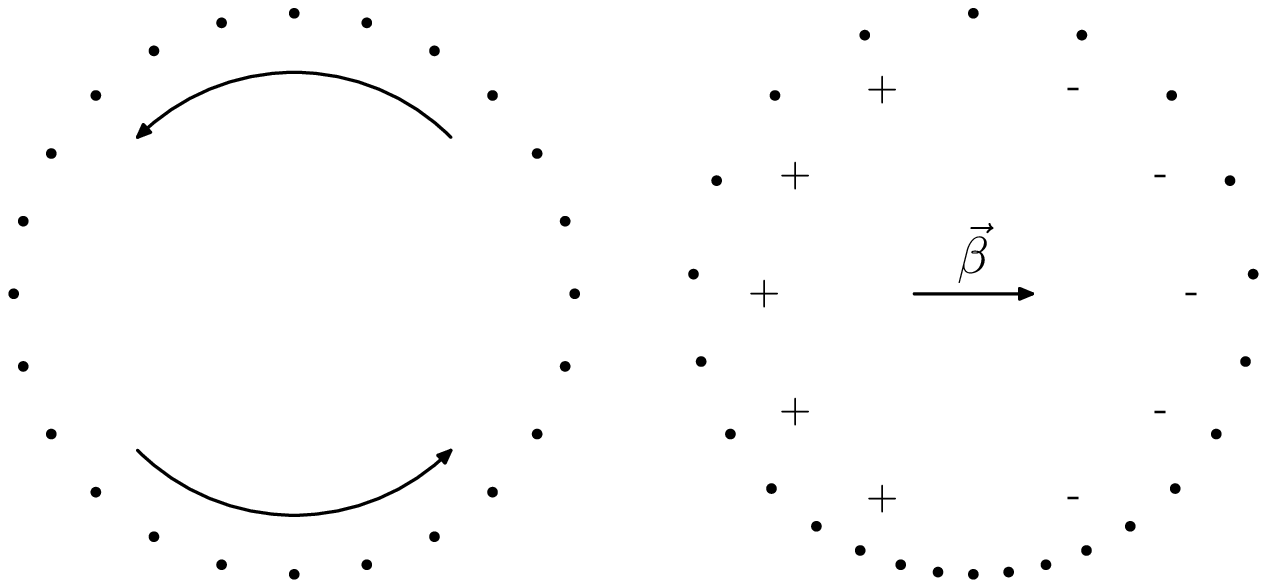}
\caption{Orbiting particle system at rest (left), showing uniform density around the orbit,
and boosted (right), showing varying density. Arrows at left indicate
the orbital velocity. Symbols $+,-$ at right indicate the sign of $\delta t$, Eq.8.}
\label{fig:boost}
\end{figure}

When the system (i.e., the nucleus) is at rest, the orbital momenta of the charges cancel and 
add nothing to the total momentum. In Newtonian physics this would also be
true of the moving system, but in a relativistic theory it need not be. 
We consider boosting the system to a small velocity $\vec{\beta}$, where $\vec{\beta}$
lies in the plane of the orbits (Fig.\ref{fig:boost}). The orbits remain
circular to first order, since length contraction starts at $\mathcal{O}(\beta^2)$, 
but they receive two types of first-order correction. 

First, the orbital velocities are altered due to the relativistic addition-of-velocity formula.  
This change does not, however, lead to any momentum discrepancy, since it corresponds to taking a snapshot
of the particle momenta at a fixed time and then Lorentz-transforming them all. The resulting total
$4$-momentum is just the Lorentz transform of the original total $4$-momentum, with no extra
discrepancy; in other words, this part gives rise to the analog of Eq.(\ref{emlor}) for the orbiting particles.

Transforming all the momenta in a fixed-time snapshot does not, however, give
an accurate picture of the moving system, since the definition of fixed time differs
between frames. Because the orbiting particles are continually accelerating, this difference
can translate into an extra contribution to the total momentum. 
More concretely, the varying orbital velocities of the particles also imply
a varying density, which creates net momentum. 

It is easiest to calculate the momentum discrepancy directly in terms of the 
internal force $\vec{F}(\vec{r})$ and the relativistic simultaneity change. 
The boosted trajectories are obtained, to first order, by substituting the coordinate change
\begin{align}
\label{lorentz}
\vec{x} &\rightarrow \vec{x}^\prime = \vec{x} - \vec{\beta} t
\nonumber\\
t &\rightarrow t^\prime = t - \vec{\beta}\cdot\vec{x}
\end{align}
into the rest-frame trajectories. The momentum discrepancy arises from the simultaneity shift,
\begin{align}
\label{simshift}
\delta t = -\vec{\beta}\cdot\vec{x},
\end{align}
 which effectively shifts the particles forwards or backwards
in their trajectories by $\delta t$, giving rise to the density shifts
shown schematically in Fig.\ref{fig:boost}. During time $\delta t$, a particle
at location $\vec{r}$ changes its momentum by $\vec{F}(\vec{r}) \delta t$,
hence the overall momentum discrepancy is
\begin{align}
\label{mattdisc}
\delta \vec{P} &= -\sum_{charges}\, (\vec{\beta}\cdot\vec{x})\vec{F}
\nonumber\\
 &= -\int d^3x\, (\vec{\beta}\cdot\vec{x})\rho\vec{E}.
\end{align}
In the last line, the equation has been written in more general form as an
integral against the orbital charge distribution $\rho$, anticipating calculations to follow.

For the circular orbits of the model, $\delta \vec{P}$ is nonzero since both 
$\vec{\beta}\cdot\vec{x}$ and $\vec{E}$ are odd under reflection through the origin; 
hence, the orbital momentum develops a discrepancy, and the next 
step is to compare this to the EM momentum discrepancy,
Eq.(\ref{discP}). For this we first need to rewrite Eq.(\ref{mattdisc}) to include the full charge distribution,
i.e. the nuclear charge as well, since Eq.(\ref{discP}) includes the E-fields arising from 
all charges. We define $\rho_1,\rho_2$ to be the orbital and
nuclear charge densities, respectively, and $\vec{E}_1, \vec{E}_2$ to be the respective
electric fields resulting from these charges. 
We can then write
\begin{align}
\label{mattdisc2}
\delta \vec{P}  &= -\int d^3x\, (\vec{\beta}\cdot\vec{x})
	(\rho_1\vec{E}_2 + \rho_2\vec{E}_1)
\end{align}
where the first term is just Eq.(\ref{mattdisc}), while the second term vanishes since
$\vec{E}_1$ is zero at the center of the orbit, where $\rho_2$ is concentrated.

Since the B-field is negligible in the system at rest, the fields $\vec{E}_i$ are derived 
from scalar potentials, $\vec{E}_i = -\vec{\nabla}\phi_i$, and one has also
$\nabla^2 \phi_i = -\rho_i$. 
Inserting the scalar potentials and integrating by parts several times, one arrives at
\begin{align}
\label{mattdisc3}
\delta \vec{P}  &= \int d^3x\, \left(-\vec{\beta}(\vec{E}_1\cdot\vec{E}_2)
			+ \vec{E}_1(\vec{\beta}\cdot\vec{E}_2) + \vec{E}_2(\vec{\beta}\cdot\vec{E}_1)\right).
\end{align}
(This result is also shown through a slightly more general method in Sec.\ref{sec:emfull}.)

This may now be compared to the EM momentum discrepancy, Eq.(\ref{discP}), with $\vec{E} = \vec{E}_1 + \vec{E}_2$,
and one sees that the matter discrepancy cancels the 
``interaction'' portion of the EM momentum discrepancy, i.e., the terms containing both $\vec{E}_1$ and
$\vec{E}_2$.\cite{griff}   It does not cancel the ``self-energy'' terms which are quadratic in $\vec{E}_1$ or
$\vec{E}_2$ alone; this makes sense because the self-energies are already implicitly included in the 
particle masses of the classical model. We note also that it is the interaction 
energy which is relevant for understanding mass differentials between different configurations
of the same charges, e.g. a capacitor with varying plate separation.

Eq.(\ref{mattdisc3}) and its comparison to the EM momentum discrepancy are, intuitively, 
the main results, but the classical model is too far removed from the true atomic theory
to be fully convincing. In addition to many other well-known deficiencies, one may doubt 
that the orbiting system would, in fact, attain the same state through 
physical acceleration which is found by Lorentz transformation, because the orbits are
largely unconstrained and could change in many ways. This doubt is considerably
reduced in the wave theory of atomic orbitals due to the discreteness of the solutions,
since a slow acceleration will not typically cause a jump between different solutions.\cite{shankar}
For these reasons, we recalculate the momentum discrepancy in the following section using the
Dirac formalism.

\section{\label{sec:dirac}Momentum discrepancy in the Dirac theory}

For a Dirac particle of charge $q$ coupled to an EM field $(\phi, \vec{A})$,  
the Hamiltonian and momentum operators are (see Appendix and ref. 16)
\begin{align}
\label{diracops}
\hat{H} = \hat{\alpha}^i\hat{p}_i + \hat{\beta}m + q\phi
\nonumber\\
\hat{p}^i = -i(\partial^i - iqA^i),
\end{align}
where $\hat{\alpha}^i$, $\hat{\beta}$ are the standard matrices of Dirac
($\hat{\beta}$ is unrelated to the boost velocity $\beta^i$).
We will consider just one Dirac field, and work in the single-particle
formalism, ignoring the complications of 
multi-electron states and the additional field for the nucleus;
these do not fundamentally change the calculation. 

Given a wavefunction $\psi$ which satisfies the Dirac equation, boosting
the wavefunction means substituting the coordinate changes, Eq.(\ref{lorentz}),
and applying a spinor rotation given by
${1\over 2}(\vec{\beta}\cdot\vec{\hat{\alpha}})$.\cite{greiner} Working for convenience at time $t=0$, the only coordinate
change is $\delta t = -\vec{\beta}\cdot\vec{x}$, the effect of which can be computed using
the Hamiltonian:
\begin{align}
\label{deltapsi}
\delta\psi &= -(\vec{\beta}\cdot\vec{x})\partial_0\psi + {1\over 2}(\vec{\beta}\cdot\vec{\hat{\alpha}})\psi
\nonumber\\
&= i(\vec{\beta}\cdot\vec{x})\hat{H}\psi + {1\over 2}(\vec{\beta}\cdot\vec{\hat{\alpha}})\psi.
\end{align}
The change in expected momentum is then given by
\begin{align}
\label{hdirac1}
\delta \left<P^i\right> = \delta\left(\int d^{3}x\,\psi^\dagger \hat{p}^i \psi\right)
= \int d^{3}x\,\left(\delta\psi^\dagger \hat{p}^i \psi + \psi^\dagger \hat{p}^i \delta\psi
								+ \psi^\dagger\delta\hat{p}\psi\right).
\end{align}
We choose a gauge where $\vec{A}=0$ in the rest frame (again assuming 
negligible $\vec{B}$ field), so the boosted state has $\vec{A} = \vec{\beta}\phi$
(an additional gauge transformation could accompany the boost, but there is no
reason to consider this). The change
in $\vec{A}$ implies $\delta\hat{p}^i = -q\phi\beta^i$, and we will also need to use
$[\hat{H},(\vec{\beta}\cdot\vec{x})] = -i\vec{\beta}\cdot\vec{\hat{\alpha}}$, which
cancels the spinor rotation terms. One finds
\begin{align}
\label{hdirac2}
\delta \left<P^i\right> &= \int d^{3}x\,\,\psi^\dagger\left(
		\left[-i\hat{H}(\vec{\beta}\cdot\vec{x})+ {1\over 2}(\vec{\beta}\cdot\vec{\hat{\alpha}})\right]\hat{p}^i 
		+\hat{p}^i\left[i(\vec{\beta}\cdot\vec{x})\hat{H}+ {1\over 2}(\vec{\beta}\cdot\vec{\hat{\alpha}})\right] 
		   -q\phi\beta^i
		   \right)\psi
\nonumber
\\
&= \int d^{3}x\,\,  \left(-i(\vec{\beta}\cdot\vec{x})\psi^\dagger[\hat{H},\hat{p}^i]\psi 
								+i\psi^\dagger[\hat{p}^i,\vec{\beta}\cdot\vec{x}]\hat{H}\psi
								-q\phi\beta^i\psi^\dagger\psi		\right)	
\nonumber
\\
&= \beta^i\left<\hat{H}-q\phi\right> + \int d^{3}x\, q(\vec{\beta}\cdot\vec{x})\partial^i\phi\psi^\dagger\psi 
\nonumber\\
&= \beta^i\left<\hat{H}-q\phi\right> - \int d^{3}x\, (\vec{\beta}\cdot\vec{x})\left<\rho\right>E^i. 
\end{align}
The first term is the expected (non-discrepancy) term, and contains the Dirac 
single-particle energy minus the EM interaction energy; this is the correct energy
to use in conjunction with the EM expression Eq.(\ref{HEM}), because
the latter already includes all matter interaction energies (see Appendix). 
The second term is the momentum discrepancy, confirming the result found from the classical model, Eq.(\ref{mattdisc}). 

This calculation, based as it is on a fixed background EM field $(\phi, \vec{A})$, still
does not account for self-energies, which are again implicitly included in the particle
masses. A full calculation would require quantum field theory and renormalization,
but otherwise would follow the same manipulations leading to Eq.(\ref{hdirac2}), and indeed
would lead to the same result, at lowest order. 

\section{\label{sec:emfull}The EM momentum discrepancy more generally}

The formula (Eq.\ref{discP}) for the EM momentum discrepancy involved only the EM field, 
while the matter momentum discrepancy (e.g., Eq.\ref{mattdisc}) 
involved both the EM field and the charge density. The computation outlined
above Eq.(\ref{mattdisc3}) shows that the two forms are, in fact, opposite to
each other, but it is worthwhile to show this using a slightly more general method. 

The EM energy and momentum derive from the stress-energy tensor (see Appendix)
\begin{align}
\label{emse}
{T_\mu}^\nu = -F_{\mu\alpha}F^{\nu\alpha}  + {1\over 4}{\delta_\mu}^\nu F^{\alpha\beta}F_{\alpha\beta}
\end{align}
where $F_{\mu\nu} = \partial_\mu A_\nu - \partial_\nu A_\mu$ is the
EM field strength. 
In the presence of charges, ${T_\mu}^\nu$ is not conserved but rather satisfies 
\begin{align}
\label{emcons}
\partial^\mu {T_\mu}^\nu = F^{\nu\alpha}J_\alpha
\end{align}
where $J^\mu$ is the total charge current, with $\rho = J^0$.
The EM momentum and energy Eqs.(\ref{HEM},\ref{PEM}) are given by
\begin{align}
\mathcal{P}^i_{EM} = \int d^3x\, {T_0}^i, 
\nonumber\\
\mathcal{H}_{EM} = \int d^3x\, {T_0}^0
\end{align}
and we wish to calculate the variation of  $\mathcal{P}^i_{EM}$ under the small boost
Eq.(\ref{lorentz}). This is given by the tensor index transformations
plus the time change: 
\begin{align}
\delta\mathcal{P}^i_{EM} 
= \int d^{3}x\, \{\beta^i{T_0}^0 - \beta^j{T_j}^i 
 - (\vec{\beta}\cdot\vec{x})\partial_{0}{T_0}^i\}.
\end{align}
In the last term we can substitute Eq.(\ref{emcons}) and
integrate the resulting spatial derivatives by parts, cancelling the second term
and arriving at
\begin{align}
\label{emdisc}
\delta\mathcal{P}_{EM}^i = \beta^i\mathcal{H}_{EM} 
		+ \int d^{3}x\, (\vec{\beta}\cdot\vec{x})F^{i\alpha}J_\alpha
\end{align}
which, for negligible $\vec{B}$ field, and noting $E^i = F^{0i}$, reduces to
\begin{align}
\label{emdisc2}
\delta\mathcal{P}_{EM}^i = \beta^i  \mathcal{H}_{EM}
		+ \int d^{3}x\, (\vec{\beta}\cdot\vec{x})\rho E^i 
\end{align}
showing a discrepancy exactly opposite to the matter momentum discrepancy of Eqs.(\ref{mattdisc},\ref{hdirac2}).

\section{\label{sec:conc}Conclusion}

The apparent discrepancies in back-reaction calculations involving classical charge
distributions are thus accounted for by compensating discrepancies within the materials 
holding the charges. This is precisely the mechanism proposed by Poincar\'{e} many
years ago; our purpose has been to describe it 
more concretely in terms of the underlying matter theories which are now known to hold. 

More specifically, the external electric field which is sourced by the classical
charges (e.g., charges on the capacitor plates) develops a momentum discrepancy 
Eq.(\ref{discP}), which seems puzzling because the classical charges are fixed in place
and cannot store additional momentum. However, Eq.(\ref{mattdisc3}) shows that
the full EM momentum discrepancy, both external and internal, is cancelled by the corresponding
matter discrepancies, when all charges are included. The classical charges do not
contribute to this because (noting Eq.\ref{mattdisc}) the total E-field at their locations
must vanish in order for the charges to be static. 

Discrepancies of the ``4/3'' variety are seen to be inevitable 
when boosting a composite relativistic system, because the boost alters the simultaneity surface
differently in different directions, changing the way momentum is divided
between the subsystems. Another way to understand the discrepancies is to note 
that the back-reaction from one subsystem onto another arises from a local interaction,
which has no way of accounting for the total energy contained in the extended region
occupied by the subsystem. This was
not an issue in Newtonian physics because interactions in that paradigm
are not mediated by local fields, hence do not cause energy and momentum 
to be dispersed over an extended spatial region to begin with. 

A more formal way to see the orgin of discrepancies is to
observe that the interaction Lagrangian $\mathcal{L}_I$ typically contains
no time derivatives (cf. Appendix), hence does not contribute to the momentum density ${T_0}^i$,
but the energy ${T_0}^0$ does contain $-\mathcal{L}_I$. The
boosted momentum must then contain a term $-\vec{\beta}\mathcal{L}_I$, and this can only be
produced through discrepancies in the subsystem momenta.

\section*{Appendix: Derivation of Energy-Momentum Expressions}

For completeness in understanding Sections \ref{sec:dirac} and \ref{sec:emfull},
we show the derivation of fundamental energy- and momentum-related 
quantities for the EM and Dirac fields. The action for an EM field coupled
to fermions of charge $q$ is given by $\mathcal{L} = \mathcal{L}_{EM} + \mathcal{L}_\psi$, where\cite{burgess}
\begin{align}
\mathcal{L}_{EM} &= -{1\over 4}F^{\mu\nu}F_{\mu\nu}
\nonumber\\
\mathcal{L}_\psi &= -\bar{\psi}\gamma^\mu(\partial_\mu - iqA_\mu)\psi - m\bar{\psi}\psi
\end{align}
where $F_{\mu\nu} = \partial_\mu A_\nu - \partial_\nu A_\mu$. The EM field satisfies the equation
\begin{align}
\label{emeq}
\partial_\mu F^{\mu\nu} = -J^\nu
\end{align}
where $J^\nu$ is the charge current $iq\bar{\psi}\gamma^\nu \psi$. We note that 
in the mostly-plus metric convention one has $\bar{\psi} = i\psi^\dagger\gamma^0$,
giving $J^0 = -q\psi^{\dagger}(\gamma^0)^2\psi = q\psi^\dagger \psi$,
which is the expected form of the charge density. 

The matter equation of motion is
\begin{align}
\gamma^\mu(\partial_\mu -iqA_\mu)\psi + m\psi = 0
\end{align}
from which one derives
\begin{align}
\partial_0\psi &= \gamma^0\gamma^i(\partial_i - iqA_i)\psi + m\gamma^0\psi + iqA^0\psi
\nonumber\\
&= -i\hat{H}\psi
\end{align}
where the single-particle Dirac Hamiltonian $\hat{H}$ is as in Eq.(\ref{diracops}), and
we note the mostly-plus definitions $\hat{\beta} = i\gamma^0$
and $\hat{\alpha}^i = -\gamma^0\gamma^i$, as well as $\phi = A^0$.  

The canonical stress-energy (SE) tensors are given by\cite{jackson}
\begin{align}
\label{emdef1}
{(T_{EM})^{\mu}}_{\nu} &= {\delta\mathcal{L}_{EM}\over\delta\partial_{\mu}A^\alpha}\partial_{\nu}A^\alpha
- {\delta^\mu}_{\nu}\mathcal{L}_{EM}
\nonumber\\
&= -F^{\mu\alpha}\partial_\nu A_\alpha - {\delta^\mu}_\nu\mathcal{L}_{EM}
\end{align}
and
\begin{align}
\label{emdef2}
{(T_\psi)^{\mu}}_{\nu} &= {\delta\mathcal{L}_{\psi}\over\delta\partial_{\mu}\psi^\sigma}\partial_{\nu}\psi^\sigma
- {\delta^\mu}_{\nu}\mathcal{L}_{\psi}
\nonumber\\
&= -\bar{\psi}\gamma^\mu\partial_\nu\psi - {\delta^\mu}_{\nu}\mathcal{L}_{\psi}.
\end{align}
The total energy and momentum of the system are then defined by
\begin{align}
\label{emtot}
\mathcal{H} &= \int d^3x\, \left\{{(T_{EM})_0}^0 + {(T_\psi)_0}^0\right\}, 
\nonumber\\
\mathcal{P}^i &= \int d^3x\, \left\{{(T_{EM})_0}^i + {(T_\psi)_0}^i\right\}.
\end{align}
These overall quantities are gauge invariant since they are derived from 
a gauge-invariant Lagrangian, but the separate EM and matter tensors are
not gauge invariant, and using ${(T_{EM})_{\mu}}^{\nu}$ one does not obtain
the familiar expressions Eqs.(\ref{HEM},\ref{PEM}). To avoid this inconvenience
one usually redefines the EM tensor to the gauge-invariant form
\begin{align}
\label{emse}
{(\bar{T}_{EM})_\mu}^\nu = -F_{\mu\alpha}F^{\nu\alpha}  - {\delta_\mu}^\nu\mathcal{L}_{EM},
\end{align}
which gives the normal expressions Eqs.(\ref{HEM},\ref{PEM}). 
This change corresponds to adding an extra term 
\begin{align}
\label{extrat}
\Delta{T_\mu}^\nu = {F_\mu}^\alpha\partial_\alpha A^\nu
\end{align}
to the canonical EM tensor, and the same must then be subtracted from the matter tensor. 
Both SE tensors are then separately gauge invariant. We note that the
overall SE tensor is not symmetric, although the EM side becomes
symmetric after the correction. The fully symmetrized SE tensor
is obtained by adding a further total divergence to the matter SE
tensor, which does not change the integrated energy and momentum.\cite{jackson} 

Hence, to verify that matter momentum discrepancies cancel those generated 
by the standard EM expressions Eqs.(\ref{HEM},\ref{PEM}), one must use
energy and momentum derived from the matter SE tensor
\begin{align}
{(\bar{T}_{\psi})_\mu}^\nu = {(T_\psi)_{\mu}}^{\nu} - {F_\mu}^\alpha\partial_\alpha A^\nu
\end{align}
which looks strange, but gives a reasonable result because
the added term can be integrated by parts and then converted to a $\psi$ term using Eq.(\ref{emeq}). 
In this way one finds energy
\begin{align}
\mathcal{H}_\psi &= \int d^3x\,\,\{{(T_\psi)_0}^0 - {F_0}^\alpha\partial_\alpha A^0\}
\nonumber\\
&= \int d^3x\,\, \{{(T_\psi)_0}^0 + J_0A^0\}
\nonumber\\
&= \int d^3x \,\, \psi^{\dagger}(\hat{H} - q\phi)\psi
\end{align}
showing the correctness of the energy used in Eq.(\ref{hdirac2}). For the
Dirac field momentum one finds similarly that the added SE term
converts derivatives $\partial_i$ to the gauge-invariant
form $\partial_i - iqA_i$, justifying the gauged momentum operator given in Eq.(\ref{diracops}). 
We note lastly that the SE tensor alteration Eq.(\ref{extrat}) does not affect
the momentum discrepancies in the two subsystems, as this term by itself has no discrepancy. 

\section{Acknowledgments}
The author gratefully acknowledges helpful discussions with and suggestions from Michael Lennek and Larry Hoffman,
as well as very valuable criticism and suggestions from two anonymous reviewers.


%merlin.mbs apsrev4-1.bst 2010-07-25 4.21a (PWD, AO, DPC) hacked
%Control: key (0)
%Control: author (8) initials jnrlst
%Control: editor formatted (1) identically to author
%Control: production of article title (-1) disabled
%Control: page (0) single
%Control: year (1) truncated
%Control: production of eprint (0) enabled
\begin{thebibliography}{0}%
\makeatletter
\providecommand \@ifxundefined [1]{%
 \@ifx{#1\undefined}
}%
\providecommand \@ifnum [1]{%
 \ifnum #1\expandafter \@firstoftwo
 \else \expandafter \@secondoftwo
 \fi
}%
\providecommand \@ifx [1]{%
 \ifx #1\expandafter \@firstoftwo
 \else \expandafter \@secondoftwo
 \fi
}%
\providecommand \natexlab [1]{#1}%
\providecommand \enquote  [1]{``#1''}%
\providecommand \bibnamefont  [1]{#1}%
\providecommand \bibfnamefont [1]{#1}%
\providecommand \citenamefont [1]{#1}%
\providecommand \href@noop [0]{\@secondoftwo}%
\providecommand \href [0]{\begingroup \@sanitize@url \@href}%
\providecommand \@href[1]{\@@startlink{#1}\@@href}%
\providecommand \@@href[1]{\endgroup#1\@@endlink}%
\providecommand \@sanitize@url [0]{\catcode `\\12\catcode `\$12\catcode
  `\&12\catcode `\#12\catcode `\^12\catcode `\_12\catcode `\%12\relax}%
\providecommand \@@startlink[1]{}%
\providecommand \@@endlink[0]{}%
\providecommand \url  [0]{\begingroup\@sanitize@url \@url }%
\providecommand \@url [1]{\endgroup\@href {#1}{\urlprefix }}%
\providecommand \urlprefix  [0]{URL }%
\providecommand \Eprint [0]{\href }%
\providecommand \doibase [0]{http://dx.doi.org/}%
\providecommand \selectlanguage [0]{\@gobble}%
\providecommand \bibinfo  [0]{\@secondoftwo}%
\providecommand \bibfield  [0]{\@secondoftwo}%
\providecommand \translation [1]{[#1]}%
\providecommand \BibitemOpen [0]{}%
\providecommand \bibitemStop [0]{}%
\providecommand \bibitemNoStop [0]{.\EOS\space}%
\providecommand \EOS [0]{\spacefactor3000\relax}%
\providecommand \BibitemShut  [1]{\csname bibitem#1\endcsname}%
\let\auto@bib@innerbib\@empty
%</preamble>
\end{thebibliography}%


\begin{thebibliography}{5}

\bibitem{thomp} J.J. Thomson, ``On the Electric and Magnetic Effects Produced by the Motion of Electrified Bodies,''
				Philosophical Magazine {\bf 11}, 229-249 (1881).
\bibitem{griff} D.J. Griffiths and R.E. Owen, ``Mass Renormalization in Classical Electrodynamics,'' Am. J. Phys. {\bf 51}, 1120-1126 (1983).
\bibitem{feyn} R. Feynman, \textsl{Feynman Lectures on Physics}, (Addison-Wesley, 1970), Ch. II.28 

\bibitem{wil1} F. Wilczek, ``Happy Birthday, Electron,'' Sci. Am. June 2012, 24. 

\bibitem{res} D.J. Griffiths, ``Resource Letter EM-1: Electromagnetic Momentum,'' Am. J. Phys {\bf 80} (2012), 7-18, 
and references therein.
\bibitem{poinc} H. Poincar\'{e}, ``Sur la Dynamique de l'\'{E}lectron,'' Rendiconti del Circolo matematico di Palermo 
			{\bf 21}, 129-176 (1906).
\bibitem{elec}for example, P. Pearle, ``Classical Electron Models,'' in \textsl{Electromagnetism: Paths to Research}, 
		edited by D. Teplitz (Plenum, New York, 1982);  F. Rohrlich, \textsl{Classical Charged Particles}, (Addison-Wesley, 1965);
J. Frenkel, ``4/3 Problem in Classical Electrodynamics,'' Phys. Rev. E  {\bf 54}, 5859-5962 (1996). 
\bibitem{bell}J.S. Bell, ``How to Teach Special Relativity,'' reprinted in \textsl{Speakable and Unspeakable in Quantum Mechanics}, 2nd ed.  (Cambridge University Press, Cambridge, 2004), pp. 67-80.
\bibitem{brown}H.R. Brown, \textsl{Physical Relativity}, 1st ed.  (Oxford University Press, Oxford, 2005).
\bibitem{miller}D.J. Miller, ``A Constructive Approach to the Special Theory of Relativity,'' Am. J. Phys. {\bf 78}, 633-638 (2010).
\bibitem{rmr}W.M. Nelson, \textsl{Relativity Made Real}, 2nd ed.  (CreateSpace Publishing, 2013).
\bibitem{burgess}  C. Burgess and G. Moore, \textsl{The Standard Model: a Primer}, reprint edition (Cambridge University Press, 2012). A helpful translation table for metric signature conventions is found in Appendix E.

\bibitem{jackson} J.D. Jackson, \textsl{Classical Electrodynamics}, 2nd ed. (John Wiley \& Sons, New York, 1975).  
\bibitem{rindler} W. Rindler and J. Denur, ``A Simple Relativistic Paradox about Electrostatic Energy,'' Am. J. Phys {\bf 56}, 795 (1987).
\bibitem{shankar}cf. standard discussions of adiabatic perturbations in quantum mechanics, eg. R. Shankar, \textsl{Principles of Quantum Mechanics} (Plenum Press, 1980), ch. 18.

\bibitem{greiner} W. Greiner, \textsl{Relativistic Quantum Mechanics}, 3d edition (Springer, 2000).


\end{thebibliography}
\end{document}